\documentstyle[11pt,aaspp,tighten]{article}

\newcommand{\myeqn}[1]{($\!\!$~\ref{#1})}
\newcommand{\myeqna}[2]{($\!\!$~\ref{#1}-$\!\!$~\ref{#2})}

\begin{document}

\title{A General Class of Self-similar Self-gravitating Fluids}
\author{Ue-Li Pen\altaffilmark{1}}
\affil{Princeton University Observatory\\
	Princeton, NJ 08544-1001}
\altaffiltext{1}{e-mail I: upen@astro.princeton.edu}

\begin{abstract}

I present a general classification of self-similar solutions to the
equations of gravitational hydrodynamics that contain many previous
results as special cases.  For cold flows with spherical symmetry, the
solution space can be classified into several regions of behavior
similar to the Bondi solutions for steady flow.  A full description of
these solution is possible, which serves as the asymptotic limit for
the general problem.  By applying a shock jump condition, exact
general solutions can be constructed.  The isothermal case allows an
extra exact integral, and can be asymptotically analyzed in the
presence of finite pressure.  These solutions serve as analytic models
for problems such as spherical accretion for star formation, infall or
outflow of gas into galaxies, Lyman alpha cloud dynamics, etc.

Most previous self-similar results are obtained as special cases.  The
critical values for a cosmological flow with $\Omega=1$ and
$\gamma=4/3$ turn out to play a special role.
\end{abstract}


\section{Introduction}

Self-similar behavior provides an important class of unsteady
solutions to the self-gravitating fluid equations.  On one hand, many
physical problems often attain self-similar limits for a wide range of
initial conditions.  On the other hand, the self-similar properties
allow us to investigate properties of solutions in arbitrary detail,
without any of the associated difficulties of numerical hydrodynamics.
They also yield a rich class of test cases for general purpose
numerical codes of gravitational hydrodynamics.

In astrophysical scenarios gravity is often important, and in those
cases the general form of similarity solutions is very restricted.  In
many situations the asymptotic boundary consists a cold flow, in which
case the solution space can be qualitatively understood.  We can
determine the qualitative and asymptotic behaviour of all cold
collisionless self-similar flows.

This paper encompasses and generalizes the calculations of previous
papers, ranging from cosmological explosions to gravitational collapse
(\cite{ost88}, \cite{ber85b} \cite{ike83}, \cite{lem89b}, etc). The
following mathematical analysis yields more insight into the nature of
solutions.

In this paper, I will start by reviewing the self-similar ansatz.
Then I identify the critical points, which serve as the asymptotic
limit of solutions.  By assuming zero pressure, we can qualitatively
solve the problem in the velocity-density phase space.  This allows us
to construct full hydrodynamical solutions by matching a shock jump to
the cold flow.  After integrating the isothermal equations, I conclude
with some suggestions for potential applications.

\section{Equations}

The goal of this paper is to solve and classify solutions to the
spherically symmetric gravitating inviscid barytropic fluid equations:
\begin{eqnarray}
\frac{d \rho}{dt} &\equiv& \left( \frac{\partial}{\partial t}
+ v \frac{\partial}{\partial r} \right) \rho =
- \rho \frac{1}{r^2} \frac{\partial}{\partial r} (r^2 v),
\\
\frac{dv}{dt} &=& -\frac{1}{\rho} \frac{\partial p}{\partial r}
-\frac{G m}{r^2},
\\
\frac{d (p \rho^{-\gamma})}{dt} &=&  0,
\\
\frac{\partial m}{\partial r} &=& 4 \pi r^2 \rho.
\label{euler}
\end{eqnarray}

Self-similarity allows us to reduce the self-gravitating fluid
equations from partial differential equations (PDE) into ordinary
differential equations (ODE) (\cite{sed59}).

The general self-similar solution allows two dimensionful parameters.
In the case of self-gravitating fluids, one is the gravitational
constant with dimensions
\begin{equation}
[G] = M^{-1} L^3 T^{-2}.
\end{equation}
This already uniquely determines the dimensionless parametrization
of any similarity solution:
\begin{eqnarray}
v &= & \frac{r}{t} V(\lambda),
\label{vparam}\\
\rho &=& \rho_h \Omega(\lambda),
\\
p &=& \rho_h (\frac{r}{t})^2 P(\lambda),
\label{prelation}\\
m &=& \frac{4}{3} \pi r^3 \rho_h M(\lambda)
\end{eqnarray}
where
\begin{equation}
\rho_h \equiv \frac{1}{6 \pi G t^2}.
\end{equation}

The second dimensionful parameter, call it $A$, has some dimension
\begin{equation}
[A] =  \frac{T^\delta}{L}  .
\label{dima}
\end{equation}
We have required that it not contain the dimensions of mass, which one
can eliminate using the gravitational constant $G$.  Furthermore, we
scaled $A$ such that the power of length is $-1$.  Some examples
will be given shortly.

We now have determined the exact form of the dimensionless independent
variable $\lambda$
\begin{equation}
\lambda = A\frac{r}{t^\delta} \label{parameqn}.
\end{equation}
After some algebra, we obtain the equations
which any self-similar solution must satisfy:
\begin{eqnarray}
\lambda ( \frac{\Omega'(V-\delta)}{\Omega}+V') &=& 2-3V
\label{continuity}
\\
V(V-1) + \lambda V'(V-\delta) &=& -\frac{2P}{\Omega} - \lambda
\frac{P'}{\Omega} - \frac{2M}{9}
\label{dlmomentum}\\
\lambda (V-\delta)(\frac{P'}{P}-\gamma\frac{\Omega'}{\Omega}) &=& 4-2\gamma-2V
\label{pressure}
\\
M&=& \frac{3\Omega(V-\delta)}{2-3\delta}.
\label{mass}
\label{similarity}
\end{eqnarray}
Here we have integrated the mass equation, as presented by Ikeuchi,
Tomisaka and Ostriker (\cite{ike83}) hereafter referred to as ITO.  We
thus have a third order ODE with three degrees of freedom for a given
gas index $\gamma$ and scaling relation $\delta$.  The equations are scale
invariant under a change $\lambda \longrightarrow a\lambda$,
eliminating the freedom in the second dimensional parameter $A$ to
only its dimension, which determined $\delta$.  Since $A$ is a constant
of the problem, it fixes the scale if we express it as an integral over
$\lambda$.  Such an example is worked out in Ostriker and Pen (1993).
In general one can obtain this integral from the initial conditions,
as I will exemplify below.

\section{Analytic Results}

We can immediately extract some information from equations
\myeqna{continuity}{mass}.  First, we look at the critical points by
setting all derivatives with respect to $\lambda$ to zero.  This
yields the static solutions which often form the boundary conditions
of the general solution.  From \myeqn{continuity} we see that $V=2/3$,
unless $\Omega=0$, which describes a matter dominated universe.  This
is an example where the Newtonian equations yield general relativistic
exact results.  Equation \myeqn{pressure} is automatically satisfied
as long as $P=0$.  From \myeqn{dlmomentum} and \myeqn{mass} we obtain
the curious result that for $\gamma=4/3$ a solution family
$P=(1-\Omega)\Omega/9$ exists.  We recover $\Omega=1 \Rightarrow P=0$
as the expected limit.  This provides a Newtonian solution for a
universe filled with a gas supported by radiation pressure.  Note from
equation \myeqn{prelation}, however, that the pressure increases at
fixed $t$ as $r^2$ without bound, while decaying as $t^{-4}$ at fixed
$r$.  The $\gamma=4/3$ gas seems to take a special role, as was
observed by Bertschinger (\cite{ber85b}) for the special case of
uncompensated collapse.  If $\gamma \neq 4/3$, the only solution is
$\Omega=1, P=0$, corresponding to the free Hubble flow.

We can also model a universe which is filled by a massless radiation gas,
such as the early radiation dominated universe, by accounting for the
relativistic pressure.  This is achieved through addition of the
equivalent inertial and gravitational mass of the relativistic gas.
The correct equation of state has $\gamma=1$ since $p=c^2\rho/3$.  $c$
is the speed of light, which determines $A=1/c$, and thus $\delta=1$.
We require $p$ to be constant, so $P\propto\lambda^{-2}$.  The unit
value of $\delta$ automatically satisfies \myeqn{pressure}.  Write the
flat line element as $-c^2 dt^2 + dx^2$.  From the energy momentum
tensor for a relativistic gas $T^{\mu\nu}=(\rho+p/c^2)u^\mu
u^\nu+p/c^2g^{\mu\nu}$ and the conservation law $T^{\mu\nu}_{,\nu}=0$
we obtain the following modified continuity equation to first order in
$v/c$:
\begin{eqnarray}
\frac{\partial}{\partial t}\rho +
\frac{\partial}{\partial x^i} (\rho+p/c^2)v^i&=& 0 \\
\frac{\partial}{\partial t} (\rho+p/c^2)v^i + \frac{\partial}{\partial
x^j}\left((\rho+p/c^2) v^i v^j + p \delta^{ij}\right) &=& \rho\nabla\Phi
\label{relgas}
\end{eqnarray}
We have written the 4-velocity
$u^\alpha=(\sqrt{c^2-v^2}/c,v^1,v^2,v^3)/c$, and $\Phi$ is the
Newtonian potential.  \myeqn{relgas} differs
from the classical Euler equation by the appearance of $\rho+p/c^2$ in
place of $\rho$.  Then critical point now  becomes $V=1/2$, the
general relativistic expansion law for a radiation dominated universe.

Let us continue the search for critical points.  At $\lambda=\infty$
we get another solution $P=0, \Omega=0, V=0$, with
$\Omega'/\Omega=2/(\delta \lambda)$.  This would correspond to an ever
diminishing density at infinity, and might be of interest in star
formation problems.  The remaining choices are to let one of the
variables diverge, with either $V=0$ which we will study in some
detail, or $V = -\infty$ which leads to a degenerate class of
solutions.  The limit $V=\delta$ is especially curious, since that
generates divergent densities moving at finite velocities.  We will
examine each of these possibilities below.

Through a change of variables $\xi=\ln(\lambda)$, we obtain a
homogeneous set of equations which do not contain the independent
variable explicitly.  We can now ask for the general form of a
solution.  One possible boundary condition at either $\lambda=0$ or
$\lambda=\infty$ is given by the Hubble solution.  Furthermore, since
the mass equation \myeqn{mass} must be positive, we have $V<\delta$
as long as $\delta>2/3$, otherwise $V>\delta$.  The latter case
implies that the fluid speed is bounded from below.  To allow negative
velocities one needs to resort to $\delta<0$.
We have a discrete symmetry for $\lambda \longrightarrow -\lambda$.
So in principle one can construct solutions for $t<0$ which encounter a
singularity at $t=0$, similar to the scaling solution for
textures\cite{dns90}.

Now let us interpret some few choices of $\delta$.
A special choice of $4/5$ corresponding to a second
parameter with dimensions of energy.  We can express our constant $A$
in terms of the energy $E$ as
\begin{equation}
A = (G E)^{-5}
\end{equation}
which satisfies our form given in (~\ref{dima}).  This solution is
investigated in further detail in \cite{ost93}.

We can derive the $\delta=8/9$ solution as found by Bertschinger
(\cite{ber85b}) from a simple dimensional viewpoint.  He starts with a
top hat perturbation of mass excess $M_i$ defined in his terms as
\begin{equation}
M_i = \delta_i \rho_h r_i^3
\end{equation}
at a time $t_i$.  We require $G M_i/r_i=GM_i/t_i^{2/3}$ to be a constant
scale of the problem, leading us to the dimensional constant
\begin{equation}
A = \frac{t_i^{2/9}}{(G M_i)^{-1/3}}
\end{equation}
with $\delta=8/9$ as desired.  We will see below how to derive the
scaling more rigorously.

An isothermal gas is characterized by its sound speed, so $\delta=1$
with simply $A = 1/v_{sound}$.  We will return to this case in more
detail in section ~\ref{isothermal}   If the constant $A$ had dimensions
of mass it would imply $\delta=2/3$, which is forbidden by the mass
equation (~\ref{mass}), showing that self-similar solutions with
characteristic mass do not exist.

\section{Cold Flows}

Often we are interested in situations where the fluid is cold at large
distances.  \cite{lem89b} had looked at pressureless cold flows
without shell crossing in a Lagrangian frame.  We can now
qualitatively solve all such cases in a Eulerian coordinate system.

When we take $P=0$, all solutions are limited to the $\Omega-V$ phase
plane.  \myeqna{continuity}{mass} become the simple system
\begin{eqnarray}
\frac{\dot{\Omega}}{\Omega} &=& \frac{2-3V-\dot{V}}{V-\delta} \\
\frac{V (6-9\delta)}{V-\delta} &=& \frac{\dot{V}
(6-9\delta)+2\Omega}{1-V}.
\label{pressureless}
\end{eqnarray}
A dot denotes differentiation with respect to $\xi$.  The homogeneous
equation in two unknowns can be completely classified once we know the
critical points.  First we note that the physically allowed region is
given by $\Omega > 0$ and $V (3\delta-2) < \delta$.  The critical
points in this region are $P_0=(\Omega=1,V=2/3),\ \
P_1=(\Omega=0,V=0), \ \ P_2 = (\Omega=\infty,V=\infty), \ \
P_3=(\Omega=\infty, V=\delta), \ \ P_4=(\Omega=0, V=1)$.  The
last point $P_4$ moves to $P_3$ when $2/3<\delta<1$.  The points
$P_3$ and $P_4$ differ also from the other three in the sense that the
solution curve reaches those points at a finite value of the parameter
$\xi$.

The critical point $P_0$ has two asymptotics leaving and two entering,
dividing phase space into four disconnected regions.  First, consider
$\gamma \neq 4/3$ and $1>\delta > 2/3$.  The second condition follows
from equation \myeqn{mass} by requiring mass to be positive.  Figure
{}~\ref{cold1} is characteristic of such solutions.  It shows the
$\Omega-V$ solutions near the critical point for $\delta=4/5$.  Two
solutions originate from the critical Hubble flow, and two converge
onto it, dividing the phase space into four regions as follows:

I. Solutions are infinitely dense at the origin, with velocities going
infinitely negative.  One might try to interpret this in terms of cold
matter accreting onto a black hole.  Far away, density and velocity go
to zero.  These solution are very attractive to many astrophysical
scenarios involving infall.  One can see some solutions involving
positive $V$ corresponding to outflow, while below a certain limit all
solutions involve purely negative velocities.  The one remaining
parameter freedom corresponding to the slope of the trajectory ar
critical point $P_1$ gives the modeler more freedom to fit
self-similar evolution to astrophysical phenomena.  The divergences
can be prevented by introducing a hydrodynamic shock as discussed
below.

II. Velocities are always positive, starting at a finite value of
$\delta$, where the density is infinite, with the total mass $M$ going
to $0$.  Both $V$ and $\Omega$ decrease to zero at infinity.  This is
a collisionless analogy to the ITO solution of an expanding hole,
except this time through a less dense medium.  This family of
solutions will also be palatable to explosive shock continuations, and
may correspond to some real situations.

III. The matter is constrained to a shell of finite thickness, such as
the propagation of exploding shells through empty space
(\cite{ike83}).  The cumulative amount of matter $m$ diverges at the
outer boundary, since a finite conserved quantity is not allowed.
Densities diverge at the inner and outer boundaries, which both move
as $r_b \propto t^\delta$.  If interpreted as expanding shells of
matter, we are looking at a decelerating shell.  So far this only
holds for cold matter, an assumption which would not hold for
explosion powered expansion.  This and the next class may not be very
physical since they require cold, divergent masses at a finite radii.
Numerical studies show that adding a bit of pressure changes the
behaviour radically.  These solutions can be obtained by matching a
shock jump, as described below.

IV. The last region again does not seem to correspond to any common
physical condition.  The cold gas is constrained to an expanding
sphere of finite radius but infinite mass, at the interior of which we
have an accreting black hole.  Again, solutions of interest may
involve invoking finite pressure.

Especially interesting are the asymptotes leading to and from the
critical point, defining the border between these regions.  Since the
vertex point $P_0$ describes the Hubble flow, the two ``incoming''
solutions correspond to a Hubble boundary condition at spatial
infinity.  The left outgoing trajectory is the unique solution where
all intensive physical quantities are bounded through all of space.
If the universe were created in this form, astronomers observing from
the origin might be tricked into believing the universe to be a Hubble
expanding big bang, when in fact densities go to zero at large
distances, giving an illusion of a locally isotropic distribution.
The right outgoing trajectory shows a cold expanding shell of matter
with finite density in the center, diverging at some finite radius.
Its existence is more a curiosity than a realistic astrophysical
scenerio, showing the range of behaviors that similarity allows.

We can quantify the solutions near this isolated critical point.
When we expand \myeqn{pressureless} to linear order for small
perturbations around $P_1$, the linear equations have eigenvalues
\begin{equation}
l_1 = \frac{3}{3\delta-2}, \ \ \ \ \ l_2 = \frac{-2}{3\delta-2}
\end{equation}
and eigenvectors
\begin{eqnarray}
v_1 &=& \left( \begin{array}{c}
	9 \delta -3 \\
		3 \delta -2
	     \end{array}
	\right)
\nonumber \\
v_2 &=& \left( \begin{array}{c}
	24 - 27 \delta \\
		6 \delta - 4
	     \end{array}
	\right)
{}.
\end{eqnarray}
The fact that the eigenvalues are real and have opposite sign
classifies this as a hyperbolic critical point, which indeed divides
phase space.  The geometric interpretation of these vectors is the
asymptotic tangent to the trajectories at $P_1$, as can be seen in
figures ~\ref{cold1}, ~\ref{cold2} and ~\ref{iso1}.  The eigenvalue
determines the direction of the arrow.

These vectors describe solutions connecting to the Hubble flow at
infinity for the negative eigenvalue, and Hubble flow at the origin
for the positive case.  For the first case, the asymptotic behaviour
of $\Omega - 1$ and $V-2/3$ goes as $\lambda^{2/(3\delta-2)}$.

An application of this analysis is that Bertschinger's infall solution
of $\delta = 8/9$ can be obtained by requiring $\Omega$ to be
monotonic and asymptotically approaching 1.  The statement corresponds
to requring a logarithmically asymptotic Hubble flow. The first
component of the second eigenvector $v_2$ must then be zero, leading
to the correct scaling.

Other scalings of interest include $\delta = 4/5$ for energy
conservation, which was investigated in further detail by
\cite{ost93}.  The isothermal case reduces to the cold flow for
sufficiently large $\lambda$, as we will show below.  This makes
Figure ~\ref{iso1} the correct solution in that limit.  We note a
qualitative change in the solutions, that the upper left corner has
actually become a valid stable asymptote.

The solution space bifurcates at $\delta=1, \ 2/3$ and $0$.  The first
case we consider in section ~\ref{isothermal}, and the last we see
illustrated in figure ~\ref{cold2}.  The bottom domain boundary is at
$V=\delta=-1/2$, so this graph includes the critical point $P_1$.
Note that the direction of the line connecting $P_1$ to $P_0$ has
reversed.

\section{Shock Jump}

We now consider solutions which actually interact hydrodynamically.
Typically this will occur when a hot gas shocks through the
cold environment, or cold matter passes through an accretion shock
onto a collapsed object.  When one region of the solution (or initial
condition) is very cold compared to the kinetic or thermal energy of
another region, the previous analysis will accurately describe the
cold domain.  The hot region is matched onto it using a strong shock
discontinuity.

We could introduce a shock jump at an arbitrary point in phase space.
What happens, however, is that such solutions either diverge, as shown
for the critical point by \cite{ike83}, or else collapse with very
divergent dynamical values.  A critical line exists in phase space,
balanced between collapse and void, which satisfy the physical
requirement of zero velocities at the origin.  For special cases of
$\delta=4/5,8/9$ these solutions have previously been found
(\cite{ost93}, and \cite{ber85b}).

Let us examine the infall solutions.  We require matter to actually
decelerate and come to a stop as it falls in.  At large distances, the
matter is taken to be cold, so that the analysis of the previous
section holds.  As we follow a solution from spatial infinity along
the line of decreasing $\lambda$, we will encounter a strong accretion
shock given by the Rankine Hugionot relations
\begin{eqnarray}
V_s &=& \delta + \frac{\gamma-1}{\gamma+1} (V_0-\delta)
\nonumber \\
\Omega_s &=& \Omega_0 \frac{\gamma+1}{\gamma-1}
\nonumber \\
P_s &=& \frac{2 \Omega_0 (V_0-\delta)^2}{\gamma+1}
\nonumber \\
M_s &=& M_0.
\label{rankine}
\end{eqnarray}
Here a subscript of $s$ denotes the postshock values and $0$ the preshock
condition.

This jump will move the curve up and to the left in figure
{}~\ref{cold1}.  The shock could occur anywhere along the solution
curve.  We will break this freedom by requiring the constraint
$V(0)=0$.  Since the problem is scale invariant in $\lambda$, the
shock jump location is not characterized by $\xi$ but by its location
in the $\Omega -V$ plane.  In order to fix the parametrization one
needs to fix the definition of the dimensional constant in terms of an
integral over the hydrodynamic quantities.  After the shock, solutions
may cross in their $\Omega -V$ projection since the pressure $P\ne 0$.
Figure ~\ref{jump} shows such trajectories corresponding to
$\delta=4/5, \gamma=5/3$, connecting to the Hubble flow or a
stationary medium respectively.  We note that solutions exist not only
for collapsing gas spheres, but also for expanding accreting ones, as
the dotted line above $V=0$ represents.

This approach reproduces the previously known solutions, and includes
generalizations for trajectories which start with stationary media and
low densities at spatial infinity.  Solutions exist for a range in
$\delta$, corresponding to various scalings as discussed earlier.

If we drop the requirement of bounded velocity at the origin, e.g. in
accretion problems onto a black hole, the shock jump location is no
longer uniquely determined.  A large family of solution exists in this
case.  Another way of increasing the size of the solution space is
to introduce several shock jumps.

The degeneracy of the solutions exists again for explosions.  It
relates to the fact that the self similar solutions remember one more
parameter from the initial conditions, corresponding to the position of
the shock jump.

\section{Isothermal Solutions}
\label{isothermal}

The existence of a characteristic velocity, such as in the case of an
isothermal flow, would lead to $\delta = 1$. In this case $\gamma=1$
and the pressure equation \myeqn{similarity} is integrable as $P = c_0
\Omega / \lambda$ for an arbitrary constant $c_0$.  While the
dimensionless parametrization \myeqn{vparam} seems rather unusual, a
scaling in terms of the sound speed e.g. $v=c V$ yield the same
equations with a change of variables $V \rightarrow V/\lambda$.  But
if we nevertheless use $G$ as the first dimensionful constant, we can
integrate the pressure equation \myeqn{pressure} to yield a new Euler
equation
\begin{equation}
V(V-1) +  \dot{V}(V-\delta) = \frac{e^{-\xi}}{2} \left(
\frac{\dot{\Omega}}{\Omega} - 3\right) - \frac{2M}{9} \ .
\label{isotherm}
\end{equation}

In terms of the autonomous equations of \myeqn{similarity} in the
variable $\xi$, the pressure term
$\exp(-\xi)(\dot{\Omega}/\Omega-3)/2$ in \myeqn{isotherm} will decay
exponentially as $\xi \rightarrow \infty$.  For sufficiently large
$\xi$, we thus recover the pressureless case, which we had
investigated earlier.  A change in behavior occurs, however, for the
critical point $V=1, \Omega=0$: it is now a stable asymptotic limit,
so solutions of this kind exist.

An isothermal shock introduces two more free parameters: the location
and strength of the shock, both of which are arbitrary.  The strong
shock conditions (~\ref{rankine}) are no longer uniquely determined,
but we nevertheless expect the same qualitative behaviors as in the
adiabatic shock, since the have seen the isothermal case to be
identical to the general solution with $\gamma = 1$.






\section{Conclusions}

We now have an improved understanding of the general behaviour of
self-similar solutions of self-gravitating fluids.  The solution space
is of managable size, and all past solutions fit into the new
framework. A class of new solutions has also been found under the
ansatz that the flow be cold at large distances.  The self-similar
motion for cold gas, which is equivalent to a pressureless fluid
without shell crossing, has been solved, with some surprising
asympotic behaviour in its $\Omega-V$ phase space near critical
points.

Since the similarity equations were derived from the gravitating Euler
equations, any similarity solution must be a valid physical scenario
under the assumptions of a polytropic inviscid gas.  Since we now know
the possible solution space, we can compare astrophysical situations
with the given templates and decide on the applicability of the
self-similar ansatz.

\section{Acknowledgements}

I wish to thank Xiaolan Huang and Professor J.P. Ostriker for helpful
discussions.  This work was supported in part by the NSF.


\clearpage
\begin{center}
{\Large \bf Figure Captions}
\end{center}

\begin{figure}
\plotone{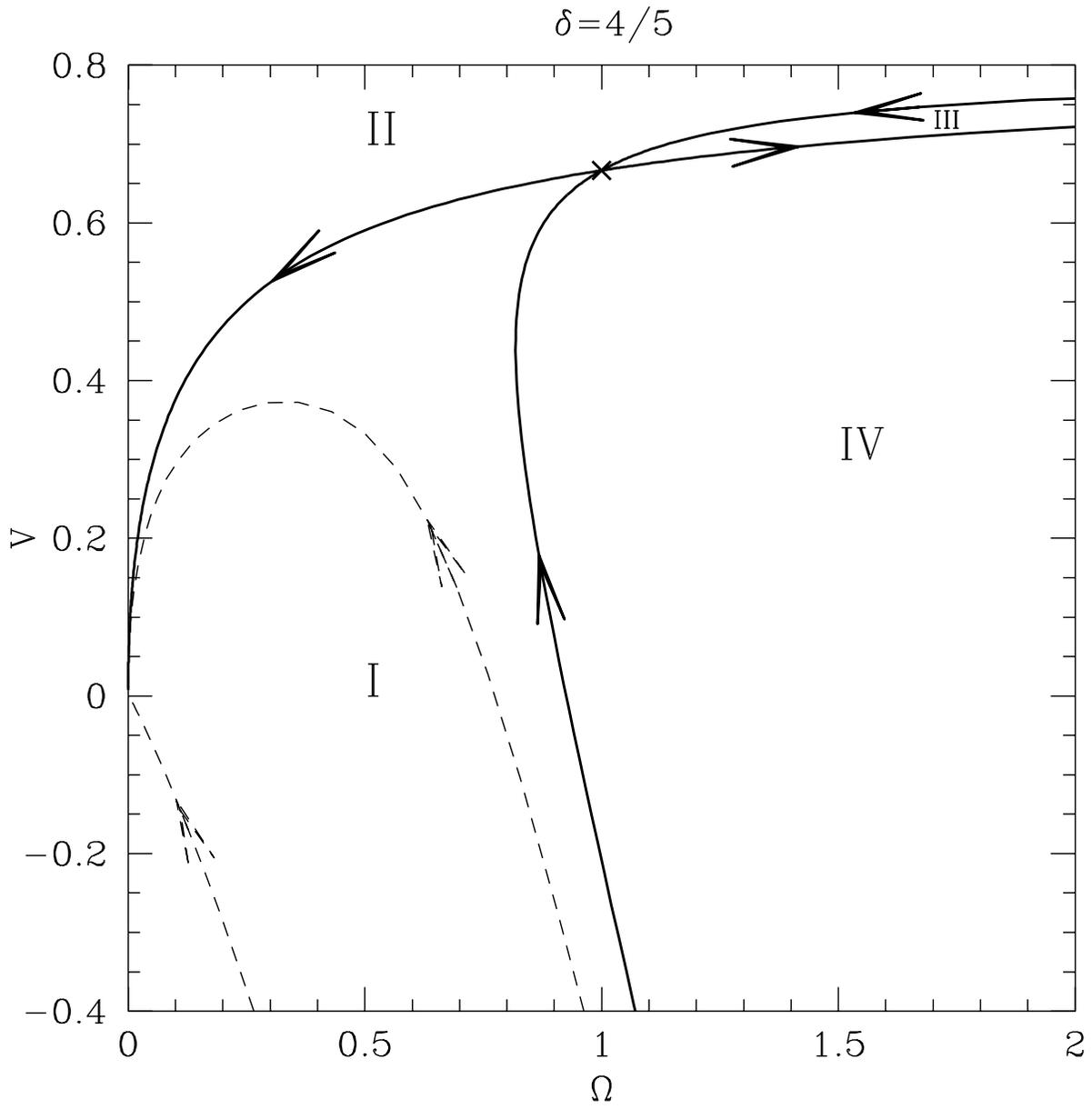}
\caption{Cold flow solution for $\delta=4/5$}
\label{cold1}
\end{figure}

Figure ~\ref{cold1}: Cold flow solution for $\delta=4/5$. Arrows
indicate direction of increasing $\xi$.  The heavy lines describe the
four asymptotic flows with a Hubble boundary condition.

\begin{figure}
\plotone{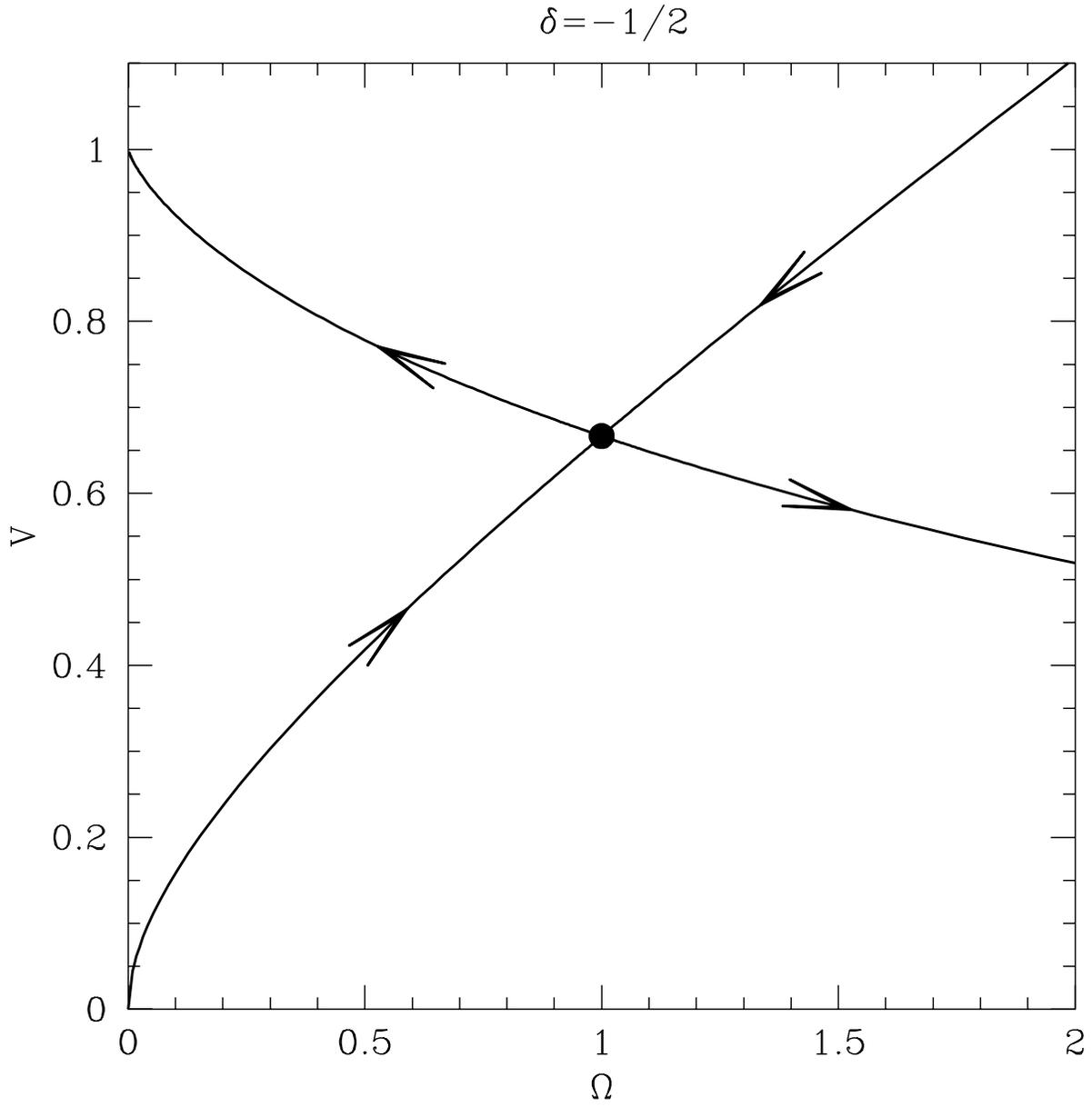}
\caption{Cold flow solution for $\delta=-1/2$}
\label{cold2}
\end{figure}

Figure ~\ref{cold2}: Cold flow solution for $\delta=-1/2$.  Here the
velocity field is bounded from below because $\delta < 2/3$.

\begin{figure}

\plotone{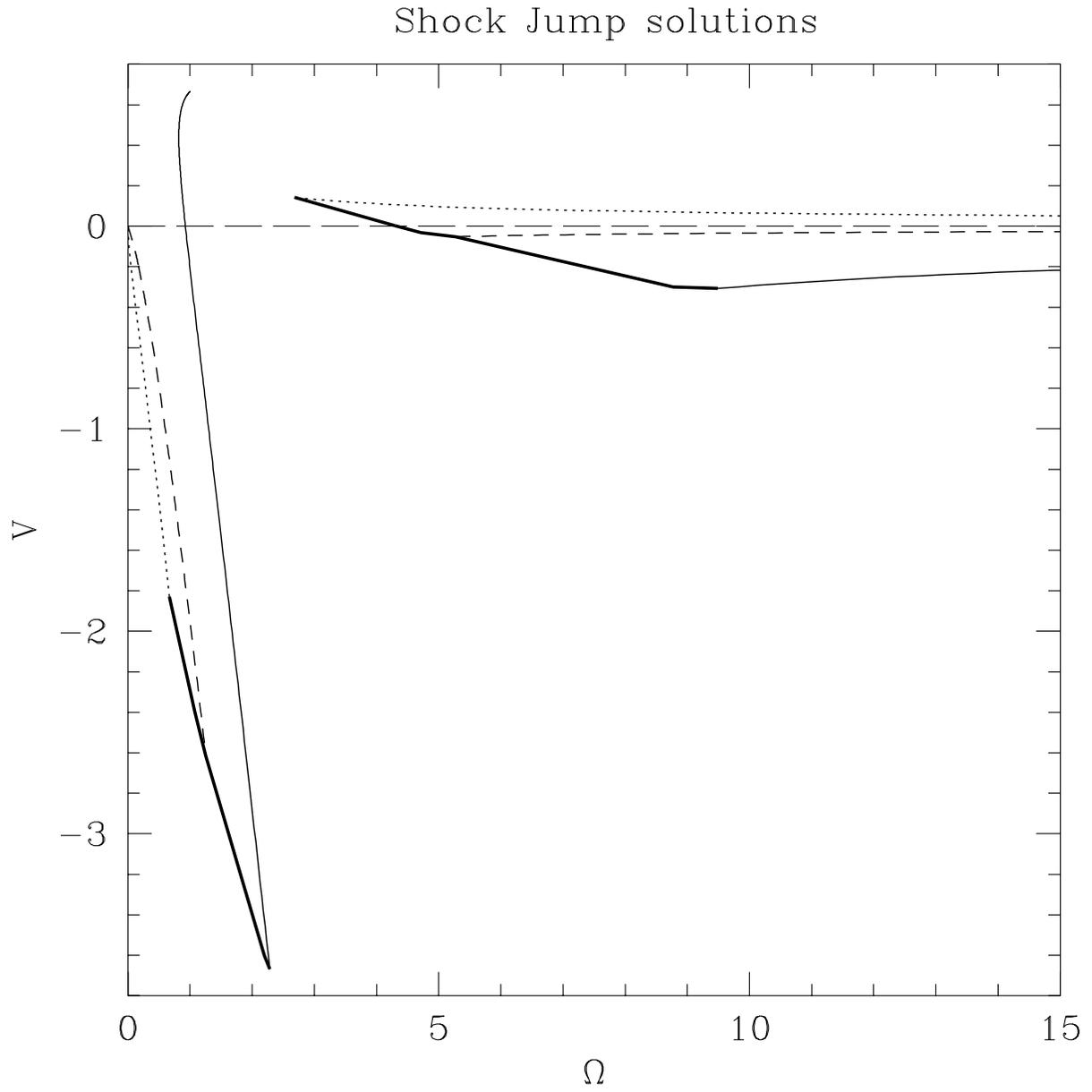}
\caption{Infall solutions for $\delta=4/5$}
\label{jump}
\end{figure}

Figyure ~\ref{jump}: Infall solutions for $\delta=4/5$ The heavy line
marks the boundary in phase space where a shock jump connects onto the
cold flow.  The solid lines represent a solution to the Hubble flow at
infinity, while the dashed line shows a solution from a stationary
medium.  The dotted line is an expanding collapse solution.

\begin{figure}

\plotone{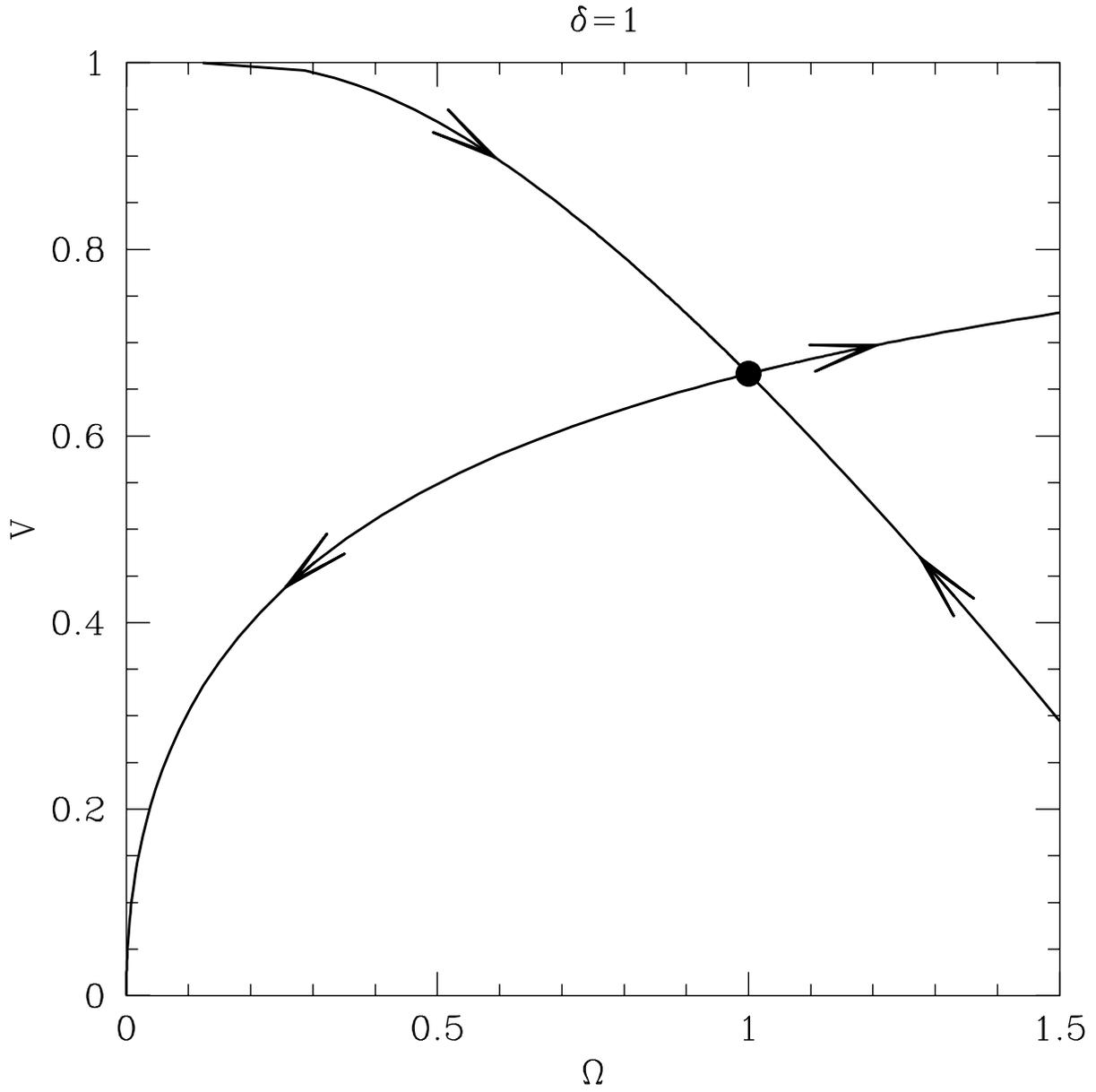}
\caption{Isothermal flow solution with $\delta=1$}
Any isothermal solution must converge to these trajectories for
sufficiently large $\lambda$.
\label{iso1}
\end{figure}

Figure ~\ref{iso1}: Isothermal flow solution with $\delta=1$.  Any
isothermal solution must converge to these trajectories for
sufficiently large $\lambda$.

\clearpage

\end{document}